# A self-organizing geometric algorithm for autonomous data partitioning

*Christopher A. Tucker*
*Cartheur Robotics*

cartheur@gmail.com

**Abstract**

A model of a geometric algorithm is introduced and methodology of its operation is presented for the dynamic partitioning of data spaces.

**Introduction**

Self-organizing algorithms are an efficient means to order complex systems [1, 2, 3]. The iterative process of drawing a geometric structure possesses the ability of self-organization when creating a connected space in the shortest path [14]. Self-organizing motifs encompass entities unto themselves and are a powerful method of representation of abstractions [4].

Geometry, its practical application, and the intellectual problems following have varied little between the ancient Greek to the Renaissance to modern times [5]. With the advent of the computer, more computationally sophisticated algorithms have been proposed [6]. Nevertheless, the central difference lies in the amount of complexity considered, in the sense of the number of inputs to the problem and the time required to solve it, parse it, and perform operations on it. In the ancient world, mathematicians dealt with a small number of inputs, calculated by hand with devices such as the straightedge and compass, and paid little attention to the complexity of the algorithm [7]. With the invention of the digital computer, studies of complexity were introduced into computational geometry [8]. For computationally elegant algorithms, simplicity is preferable to complexity.

This paper will describe a geometric algorithm drawn using the limited tools of a straightedge and drafting compass by automata. The automata, in the form of goal-seeking agents, obey a rule that dictates they draw the structure with the minimum amount of traveling while visiting each vertex and edge at least once but at most twice. The activity of the automata yields a self-organized structure of equivalent and symmetric dimensions to be leveraged as a partitioned computational data-space of a finite block size.

**The methodology of the self-organizing geometric algorithm**

The need for compiling data from high-volume sensor outputs in real-time requires autonomous data partitioning as a computational model. Using auto-sequencing memory address assignments and data counters, the number of times an agent traverses a vertex edge is important due to the demand for efficiency in the algorithm—ideally, when writing it to the partition, an iterative loop replicating space and addresses information for the transit of data. In this paper, this is the model of a geometric data processor with a level of abstraction reconciled by an algorithm using the method of exhaustion [9, 10].

The model uses a geometric address generator—demarcated by a coordinate value of three vertices—a variant on the generic address generator comprising a further generalization of direct memory access. Globally, the geometric arguments are a systolic array with a space-time diagram on a generalized Riemann manifold. In terms of stability [11], a central, fundamental vertex that is not derived is required to define the point of symmetry in the shape. This fundamental vertex serves as a point of reference to an arbitrary construct of coordinates at any location for a Cartesian grid in Euclidian space.

Practically speaking, the incoming data stream requires two program sources to control it: Flowware, the paradigm sequence by which the data is input into the array, known by the constructor agents, and configware, how the system knows where the data is located to report to a high-level language compiler to leverage data at run time. Communication of coordinates, in the form of three vertices and a logical space integer, resides in the agent automata and the application layer. This is a model of a reconfigurable data-stream machine and a model of a reconfigurable compiler. The agents create and arrange the compiler and its partitioner in reference to input from a high-level language and partitions the code into geometrically parallelizable streams suitable as a reconfigurable optimizer how locations, in the form of address by the vertices and edges, are determined for the memory utilized by a microprocessor.

Instruction-based von Neumann machines are not appropriate for such operations as execution cannot be determined *a priori* without first processing data received from sensors [12]. Rather, a reconfigurable computer or Xputer derived from a generalization of a systolic array represented by the exchange of data between the knowledge retained by its agents and the data residing within is an improvement. Only after knowledge of the data can be reconciled with where it is stored can meaningful arrangements be facilitated [13].

In terms of this paper, a set of fundamental laws define how the agents interact with the environment and properties each possess. Some of these abilities are:

1. Automata which build the shape on first iteration, traversing the structure only once,
2. Agents which possess the knowledge of their work for later search operations,
3. A limit to associated knowledge within the system so data can be exchanged without long wait times to access data and make decisions,
4. A method to pool data by simple addressing from sensor or other data-intensive hardware and networks,
5. Present the refined data to a program through an interface.

In the following sections, the construction of a geometric algorithm and analysis of its order, logic, flow, and complexity will be presented with a focus on connectivity between vertices [14].

**The walker-constructor algorithm**

Consider Fig.1, a closed spatial system consisting of *n* vertices with *m* edges, where $n = 10$ and $m = 21$. The shape is constructed by three goal-seeking agents of the form $f : P^* \to A$. The agents possess the abilities to walk, to mark, and to remember, in the scope of a collective task to construct a geometric object with the explicit rule to pass through each vertex and draw the edges only once. They possess a tool, a drafting compass, and have pins and elastic strings.

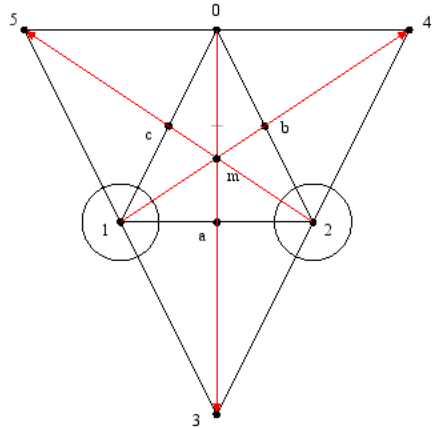

Fig.1. Walker constructor drawn by agents.

The algorithm has two components: the task and the activity. The task is to observe agents as they perform their task of obeying the rule when constructing the object. The activity begins at any vertex in a spatial grid, regardless of gauge or coordinate, in terms of the execution of the algorithm, the space is considered null. The algorithm is described as: Pick an arbitrary point on a topological manifold, call it (0,0), and run a Cartesian grid in four directions with each ray trajectory 90 degrees from its left and right nearest neighbor where $p = 2$ dimensions. Label one axis 'x' and name the other axis firstly encountered in a counterclockwise rotation 'y'.

Create three agents who are programmed identically and whose task is to walk, to mark, and to remember. Between them, they hold two elastic strings. Start each at (0,0): agent (1) walks five units in the positive y direction; agent (2, 3) walk five units in the negative y direction and walk 90 degrees from their path in opposite directions: (2) in the negative x direction 5 units and (3) in the x direction 5 units.

Remember your location on the grid.

Draw tight the first elastic string and pin to the edge points (1,2,3). This is the first polygon of three vertices. Scribe circles with the compass with a diameter of 2 units filled with quantity 'theta' centered on vertices 2 and 3.

1. Cast a ray that intersects point 'a' continuing to twice the distance walked and mark the edge point '4'. Save the distance (d4),
2. Walk in the +y direction along the string to the grid coordinate (2.5,0) and mark the point 'c',

3. Walk in the +y direction along the string to the grid coordinate (-2.5,0) and mark the point 'b',
4. Cast a ray that intersects the point 'c' continuing the ray to twice the distance traveled and mark the edge point '5'. Save the distance (d5),
5. Cast a ray that intersects the point 'b' continuing the ray to twice the distance traveled and mark the edge point '4'. Save the distance (d6),

Draw tight the second elastic string and pin to the edge points (4,5,6). This is the second polygon of three vertices. Mark the intersection of the three rays as the center of mass of the triangle centroid as point 'm'. A graphical depiction at various points in time of the execution of the algorithm is shown in Fig.2.

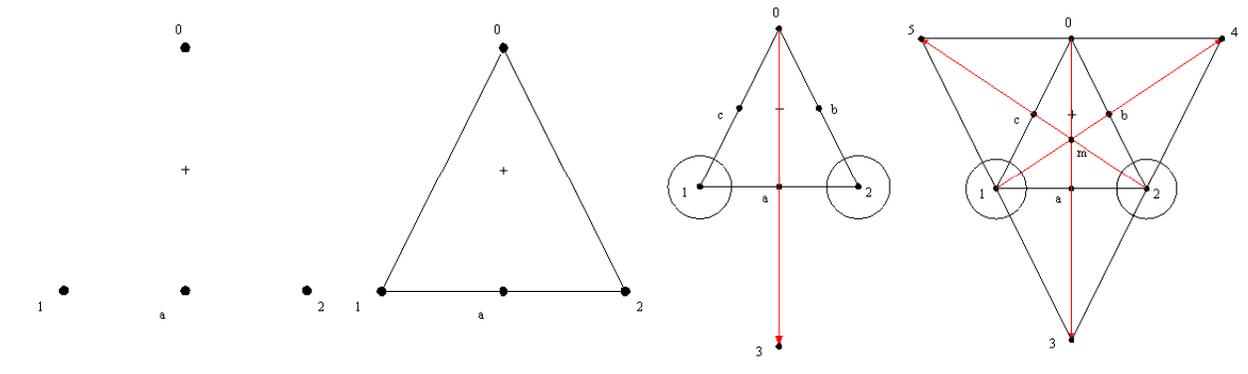

Fig.2. Walker constructor progress within algorithm execution.

The collective behavior of the agents exhibits a Hamiltonian path [15], which is NP-Complete [16]. The time to solve the solution takes a longer time than the algorithm to draw it. The walker constructor agents possess the knowledge of the structure and have the ability to perform operations on the space containing it rather than attempting such a solution in any other manner, including access of the spatial zones by application operations. One such operation is the creation of an addressable data partition, illustrated in Fig.3, that carries the knowledge to self-modify in order to posit data arriving from, for example, sensors about the external environment outside the structure.

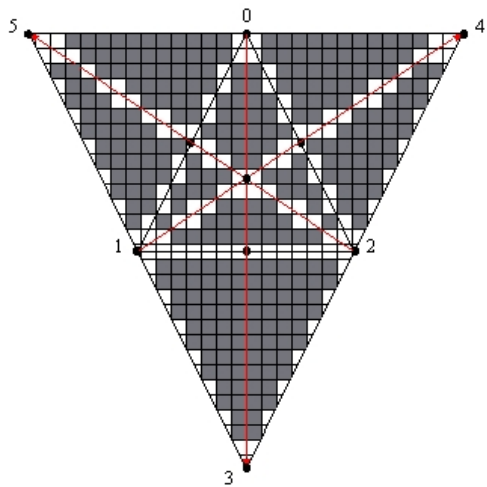

Fig.3. Data partitions following walker algorithm.

The addresses of partitions are only noted when it is a non-integral type space, that is, where a partition is not intersected by an interior edge or a ray. Illustrated in Fig.3, the partition of {5, 0, c}, has 11 blocks at the edge, 10, 8, 6, 4, and 2 toward the interior, for a total of 41 blocks. A typical address for a block would be {5, 0, c : 37}. The same number of blocks consist the partition {4, 0, b}. By symmetry, a set of partitions to the left of the ray {0, 3} have the same number of blocks as partitions to the right of the ray. The algorithm can further divide the size of the partition by manipulating the quantity of $\theta$ in its compass tool relative to the length the edge.

**Algorithm flow and complexity**

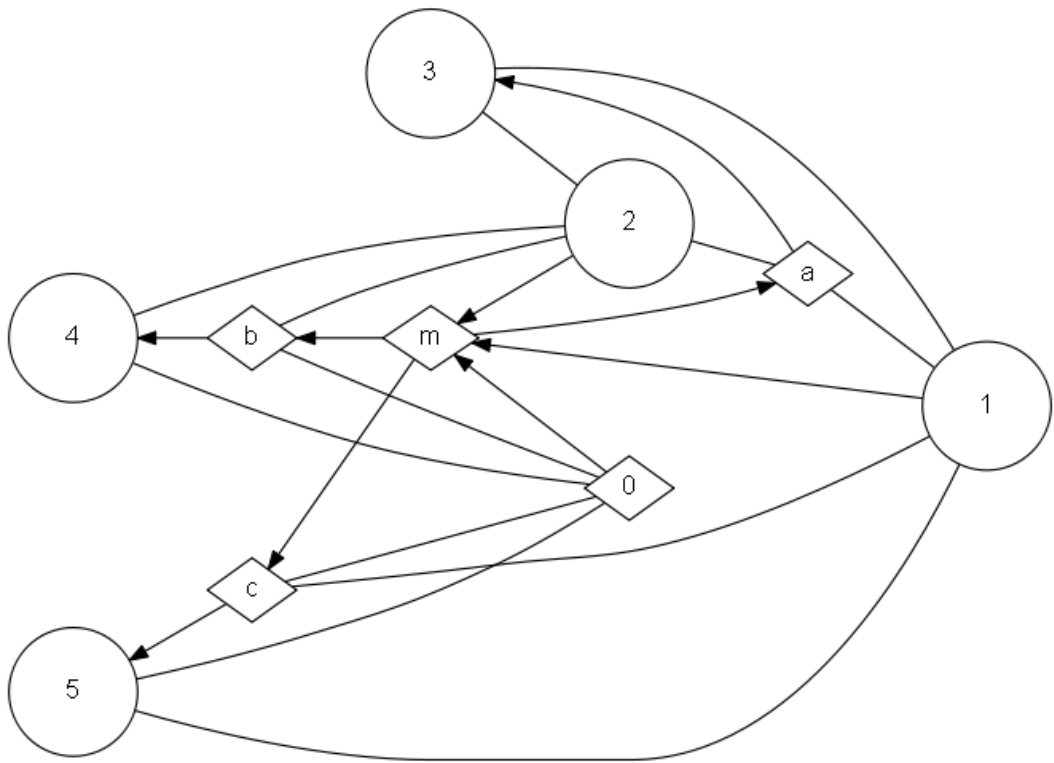

Fig.4. The adjacency list by cyclic graph.

Fig.4 is a representation of the adjacency of the vertices of the walker algorithm with both undirected and directed edges. The vertices consisting the inner triangle, $a,b,c$, with its center point, $m$, as well as the starting vertex, $0$, are represented by diamond-shaped nodes. The vertices containing the endpoints, $3,4,5$, and the compassed scribes, $1,2$, are represented by circle-shaped nodes. The vertex 'm' is the center of the diagram and shows directed edges intersecting vertices $a,b,c$, and is the subject of directed edges from $2,1,0$.

The purpose of this algorithm is to create a domination of the available paths and the knowledge of said in the shortest amount of time [17]. The computational efficiency of the algorithm is determined by an adjacency matrix, $i \times j$, where $M = [n,m]$ represents the vertices and edges. If the edges form the limit, it will take $O(1)$ time to test if an edge, $m$, is represented by the adjacency matrix as a coordinate wherein the bits in the partition are read, written, or searched in binary form. As partitions become filled, the form can expand into $p = 3$ dimensions to accommodate before subdivision. However, higher degrees of order add complexity as the as the number of edges increase. Complexity can increase very quickly if the system were to subdivide its spaces by a dimensional extension of the structure; the selection of the dimension around 'm' higher than $p = 2$ increases the weight of operations on the vertices and edges as agents are traversing more paths. The algorithm is on the maximum order of $O(n^2)$ for undiscovered states since the rule insists each vertex is completely explored and that all incident edges have been visited. Searches, either breadth-first or depth-first, performed on the algorithm are in $O(n+m)$ time.

The structure containing partitions has the following features to offer the application layer:

1. Partition addressing: stored as a location given by coordinates $\{(i,j,k):x\}$. Since the system knows how the triangles were constructed, it can retrace edges to find each of the vertices,
2. Partition resolution: the size of the blocks is arbitrary address storage where the number of spaces and sequential numbering would be determined by properties of the physical memory. Further iterations, the addition of vertex edges between vertices, can be constructed in $\Theta(d_i)$ time where $d_i$ is the degree of the $i^{th}$ vertex,
3. Expense of access operations by assigning access vertex $v_x$ to an input-output controller,
4. It describes the state values of any operation in terms of its spatial model and known dependencies of adjacency,
5. Is a closed and well-defined system with a venter represented by $m$ which interacts with all vertices as the center of mass of the object.

**Conclusion**

This paper described a geometric algorithm capable of the ability to self-organize a computational map suitable for data partitioning.

The methodology of modeling the algorithm illustrates how data manipulation systems can be architecturally arranged based on weights of information flow from a sensor array or other data-intensive service using a geometric structure with a center point of symmetry. It illustrates this by imposing stability at each of the structure's vertices and knowledge of pathways between the outer vertices and the center of mass. The alteration of the behavior of the object by making corrections to the strategy of data addressing and storage is an optimization technique.

The simplicity in the representation of drawing relative to the spaces makes this geometric algorithm appealing. Further vertices are added as the number of sub-triangles inside the numbered edges increase since at least one edge is known within the subdivided addressable space, similar to a Sierpinski carpet containing the Cantor set in two dimensions.

The author proposes the geometric algorithm illustrated in this paper is suitable as a governing algorithm, similar in form and function of DNA cores and RNA assemblers in organic systems, a set of procedural primitives for artificial life constructs in a machine.